\documentclass{kapproc} % Computer Modern font calls

\usepackage[dvips]{graphicx}

\upperandlowercase

\kluwerbib % will produce this kind of bibliography entry:

\begin{document}

%------------ article title  ------------------->>

% If you use \\'s , please supply an alternate version of the title
% in square brackets, i.e., 
%\articletitle[Communism, Sparta, and Plato]
%{COMMUNISM, SPARTA,\\ and PLATO}

\articletitle[Diffuse elliptical galaxies, the first 3D spectroscopic observations]{
DIFFUSE ELLIPTICAL GALAXIES, THE FIRST 3D SPECTROSCOPIC OBSERVATIONS}

%% optional, to supply a shorter version of the title for the running head:
%%\chaptitlerunninghead{}

\author{Igor Chilingarian}
\affil{Sternberg Astronomical Institute, Russia; 
Observatoire de Lyon, France: visiting investigator}

\author{Philippe Prugniel}
\affil{Observatoire de Lyon, Observatoire de Paris-Meudon, France}

\author{Olga Silchenko}
\affil{Sternberg Astronomical Institute, Russia}

\author{Victor Afanasiev}
\affil{Special Astrophysical Observatory, Russia}

%------ author/affiliation choices -------------->>

%% Single author

% \author{}
% \affil{}
% \email{}

%% Multiple authors, single affiliation

% \affil{}

%% Multiple authors, multiple affiliations

%------ prologue, abstract, keywords ----------->>
% optional prologue
%\prologue{<text>}{<author, year>}

% optional abstract
\begin{abstract}
Diffuse elliptical galaxies (or dwarfs) are the numerically dominant 
population in clusters like Virgo and Fornax. They carry only a few 
percent of the baryonic mass but are very sensitive to the environment; 
so they keep a fossil record of the
environmental conditions over the life of the cluster.

Recent observations of their kinematics and stellar populations cast new
insights to their evolutionary history. 
The most promising constraints come from integral field spectroscopy. We
present for the first time 3D observations of a dwarf elliptical galaxy, IC~3653.
These good signal to noise data (S/N=30) obtained with the Russian 6 meters
telescope reveal a complex kinematics reminiscent of that commonly observed
in elliptical galaxies. The disk-like feature has no morphological counterpart
in ACS data. 

Kinematical evidences for thin stellar disks confirm
and enforce the earlier imagery observations and strengthen the connection
with disk galaxies. We suggest that dE galaxies formed from small units, developing
a disk. This disk was heated or destroyed by the winds induced by the star formation, 
and the galaxy was later depleted from the
remaining gas by ram-pressure stripping when crossing the cluster.
\end{abstract}

% optional keywords
% \begin{keywords}
% Text, text...
% \end{keywords}

%------------ body of article ------------------->>
\section{Introduction}
Diffuse elliptical galaxies (dE, also called dwarf elliptical or dwarf
spheroidal galaxies) are dwarf ($10^7$ to $10^9$ $M_{\odot}$), structureless
(at least at first sight; hence their name) and diffuse: light profiles are
nearly exponential, dEs never show de Vaucouleurs profiles, the S\`ersic
exponent is normally between 0.6 and 3. They are also ubiquitous in
clusters: for instance, more than 70\%\ of known Virgo cluster members are
dEs. The dE galaxies are not well studied because they are difficult to
observe: their surface brightness, metallicity (absorption-line contrast)
and intrinsic rotation/chaotic velocities are low and therefore require long
observing time and high spectral resolution.

They were believed to be simple objects, until recent high
quality observations revealed fine structures, in particular the presence of
a spectacular and intriguing stellar spiral in IC~3328 (Jerjen et al. 2000)
and IC~783 (Barazza et al. 2002) or other structures, like embedded discs
or bars in a lot of dEs (Barazza et al. 2002).  A number of recent
papers presented also long slit spectroscopic data (Simien \& Prugniel 2002,
Pedraz et al., Geha et al 2002, 2003, de Rijcke et al. 2001, van Zee et al.
2004) revealing their diversity. The majority of dE appear rotationally 
supported, with some very noticeable exceptions of clearly anisotropic systems
(Simien \& Prugniel 2002). Some examples of kinematically decoupled cores
were found by de Rijcke et al. (2004). The stellar population also 
present a cosmic dispersion which suggests a diversity of the origin of evolutionary history
of dE (Michielsen et al. 2003). Some ionized ISM has been detected in 
some cases (Michielsen et al. 2004).

\section{Origin of dEs: possible scenarios}
The first scenario is a hierarchical collapse with a feedback of star
formation (Dekel \& Silk 1986, Mori et al. 1997). Some difficulties of this
theory existed. They were caused by the disagreement between fundamental
properties of the dEs obtained from observations and computed using common
models of that process. In particular, observed velocity dispersions for
fainter galaxies were lower than had been expected (Faber-Jackson relation).
The problem has been solved recently by taking into account effects of the
dynamical response to starburst-induced gas removal (Nagashima \& Yoshii
2004).

The other scenarios give the first role to the effects of environment.
Three main mechanisms are discussed: (1) the ram-pressure gas stripping 
of late spirals and dwarf irregulars in clusters (Mori \&
Burkert 2000) and groups (Marcolini et al. 2003); (2) the tidal harassment
due to distant and repeated encounters with other cluster members (Moore et
al. 1998); (3) the collisions between large gas-rich spirals which may 
eject massive gas tails from which a dwarf galaxy may born (tidal dwarfs) 
and evolve into a dE after fading of the young stellar population
(Duc et al. 2004). This last possibility can surely not be the main scenario 
because dE's are one order of magnitude more numerous than large galaxies
which in average have experienced one major collision in their lifetime.

It is likely that both the wind-controlled early evolution and the environment
effects combine to produce the actual state of dE. The moderate or low metallicity
of most dE indicates that the stars were formed in small galaxies subject to
the winds that prevent enrichment, but on the other hands environment effects
cannot be avoided.
Ram pressure stripping is the accepted explanation for the HI depletion of
spiral galaxies in clusters, and any gas-rich galaxy in a cluster has to
experience it. It may well explain the relation between the morphology
of dwarfs and the density of the environment (Binggeli et al. 1990,
van den Bergh 1994): low-density environments are mostly populated
by dIrr's, while dE's reside mostly in clusters where ram pressure
stripping is expected to be the most efficient. Recently, Conselice et al.
(2003) found HI in 7 dEs in the Virgo cluster, which they
interpreted as transition objects on their way in the morphological 
transformation from dIrr to dE. In a series of papers, Conselice et al.
(2001, 2002) argue that dEs have been accreted quite recently and have
formed from late type galaxies after they have crossed the
center of the cluster.

To bring additional constraints we started a survey of dE galaxies with
integral field spectroscopy. We are searching for kinematical structures, 
eventually related to morphological fine structures, and for subsequent
structures in the spatial distribution of the stellar population properties.
Evidently, both types of the structures are to be connected by a common origin.

\section{First results for IC~3653}
We are presenting here the first study of IC~3653, a dE in the
Virgo cluster, by the means of integral-field spectroscopy, with the 
MPFS (Multi-Pupil Fiber spectrograph) of the 6-m telescope ($R \sim 1500$).

The galaxy was observed under good atmospheric conditions (seeing 
$1.4^{\prime \prime}$)
in the spectral range between 4150\AA\ and 5650\AA, the total integration
time was 2.5 hours. The Voronoi adaptive binning
technique (Cappellari \& Copin, 2003) was used to keep rather high
signal-to-noise ratio over the field of view of the MPFS (16$\times$16 array
of $1^{\prime \prime} \times 1^{\prime \prime}$ elements).

To analyse the data, and in particular to attempt to constraint the
history of the stellar population we are using new technique based
on the recent evolutionary synthesis tool PEGASE.HR (Le Borgne
et al. 2004). PEGASE.HR is used to generate a grid of simple
stellar populations (SSP) covering the range of metallicities expected
for dE galaxies. These SSPs are sampled at different ages, building
finally a grid in age and metallicity. Our inversion method consists
of fitting simultaneously the line-of-sight velocity distributions (LOSVD)
and the distributions of SSPs in age and metallicity in every point
of the field of view.

The flux-calibrated SSPs were generated with the ELODIE library at a
resolution R=10000, and the first step of the analysis required to adjust
the spectral resolution to the one of the MPFS observations which varies
both with wavelength and with the position in the field of view of the
spectrograph. Since the resolution of the spectrograph is not higher than
the kinematical broadening, this resolution matching must be extremely
precise.

The principle of combining several templates in the kinematical analysis is
widely used since about 10 years in order to reduce the so-called template
miss-match. This approach, called optimal template fitting, generally
includes an additive low-order polynomial which absorbs the effects of
metallicity miss-match and of diffuse light. Therefore, the best fitting
combination cannot be interpreted in terms of stellar population. For the
present work, we have for the first time a library of templates with a
suited range of metallicities and with a high quality, hence we did not
include any additive term and attempted to interpret the resulting weight of
the SSPs as an history of the stellar population. As stated, the correction
of diffuse light is particularly critical for this analysis, and it is not
straightforward for 2D spectrographs. Our procedure was tested by
simulations which proved its accuracy.

Our MPFS data for IC~3653 being analysed by this method have revealed some
interesting properties of this Virgo dwarf elliptical. We find significant
kinematical peculiarities in the galaxy. The gradient of the line-of-sight
velocity in the central region is very steep. Velocity dispersion shows a
sharp peak slightly off-centered. It may indicate the presence of a
kinematically decoupled core as it is observed in several giant elliptical
galaxies (Franx \& Illingworth, 1998) and suggested in some other dEs
(Morelli et al. 2004). The disk-like kinematical feature has no
morphological counterpart on ACS images from the HST archive (proposal 9401
by C\^ot\'e).  The kinematical data are shown in Fig.~1 and Fig.~2

Integrated properties of the stellar population show no significant
gradients; the measurements of the Lick indices in the spectrum of the
centre of the galaxy is shown in Fig.~3. The mean stellar age derived from
these measurements is $6.5 \pm 0.8$ Gyr, the metallicity is solar ($[Fe/H] =
0.0 \pm 0.04$). The comparison of IC~3653 with the
prototypical dE NGC~205 and compact elliptical galaxy M~32  
 is shown in Tab.~1 (population data are from Trager et. al., 1998).
IC~3653 is a relatively bright object compared to the dEs
drawn from Geha et al (2003).
Though its population ressemble more to  M~32 than to  
NGC~205, its diffuse structure relates it more closely to the
sequence of dEs than to the Es to which M~32 belongs.
IC~3653 is located at the top of the dE sequence and its high metallicity
locates it on the natural mass-metallicity relation.

Our stellar population fit also indicates the solar metallicity without any
scatter and an age distribution tightly picked at 7 Gyr with marginal
indication of an intermediate age sub-population at an age of 1 Gyr with a
mass fraction of about 3\%. The difficulty of the inversion is to prove its
uniqueness, in particular because of degeneracies of the problem (between
age and metallicity, or velocity dispersion and metallicity). We have not
investigated this issue in detail yet, so the final conclusions will be
presented in a forthcoming paper; but the announced solar metallicity
appears to be robust. The kinematics in the core region is not spatially
resolved with the present observations and further observations are
required.

Evidences for kinematical structures in dE establish a link with small
spiral or dwarf irregular galaxies. Many of the dE yet studied in detail are
not genuine spheroids, they are not like big and diffuse globular clusters.
dE may generally have formed from low-mass units, which developed a disk.
The winds have controlled the enrichment and the dynamical feedback has
thickened or destroyed the disk. The ram pressure stripping, when the future
dE crossed the denses regions of the cluster finished to remove the cold gas
left, in particular in the plane of the disk, by the anisotropic winds. The
long-lived features in the stellar population can still be observed now, 5
or 10 Gyr after their formation.

\begin{figure*}
\hfil
\begin{tabular}{c c c}
 \includegraphics[width=3.5cm]{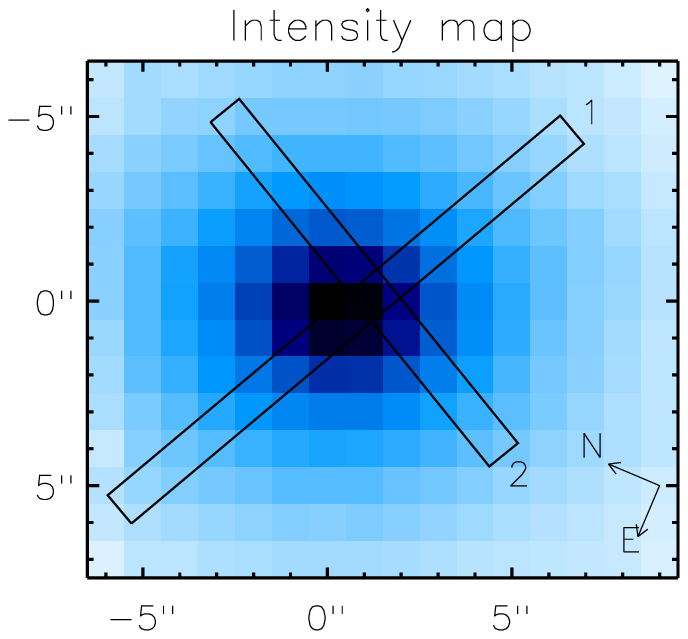} &
 \includegraphics[width=3.5cm]{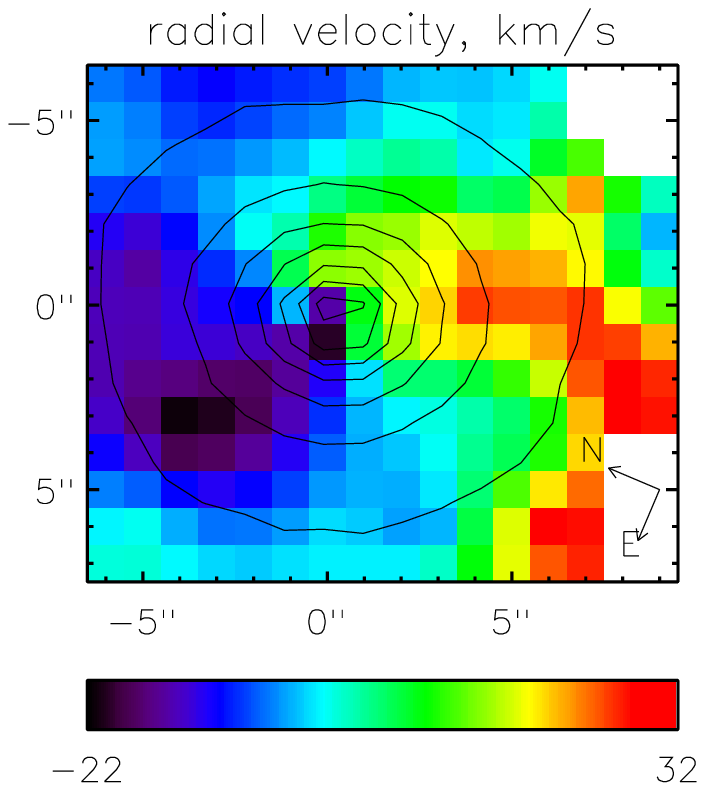} &
 \includegraphics[width=3.5cm]{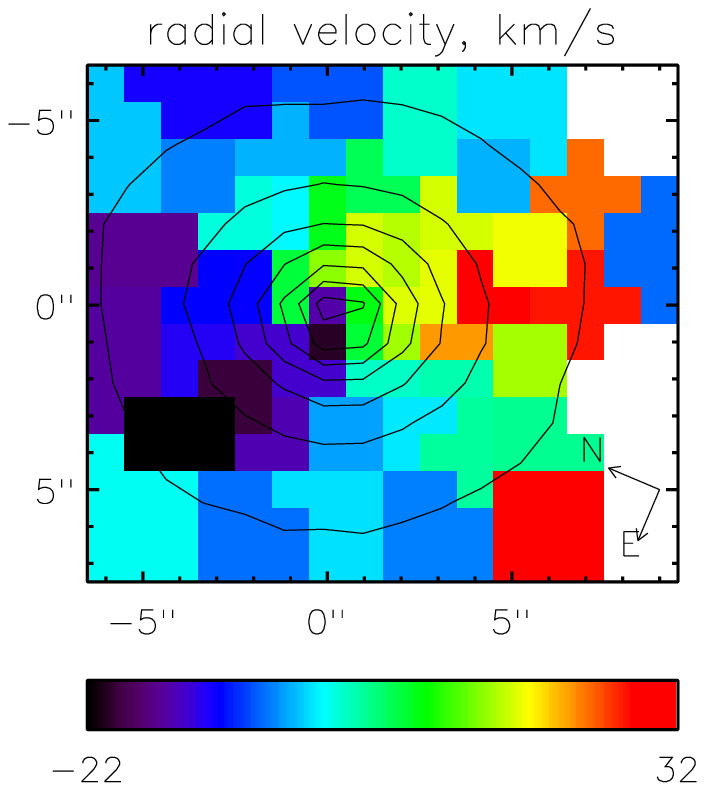} \\
 (a)&(b)&(c)\\
\end{tabular}
 \caption{The generalized view of the kinematical data.
(a) intensity map with the positions the kinematical profiles are presented
for; (b) radial velocity field, interpolated between the nodes of the
Voronoi tessellation; (c) radial velocity fields: values exactly correspond
to the Voronoi tessellae; The values with uncertainties greater than 18 km/s
are masked.}
\hfil
\label{fields}
\end{figure*}

\begin{figure*}
\hfil
\begin{tabular}{c c}
 \includegraphics[width=5.5cm]{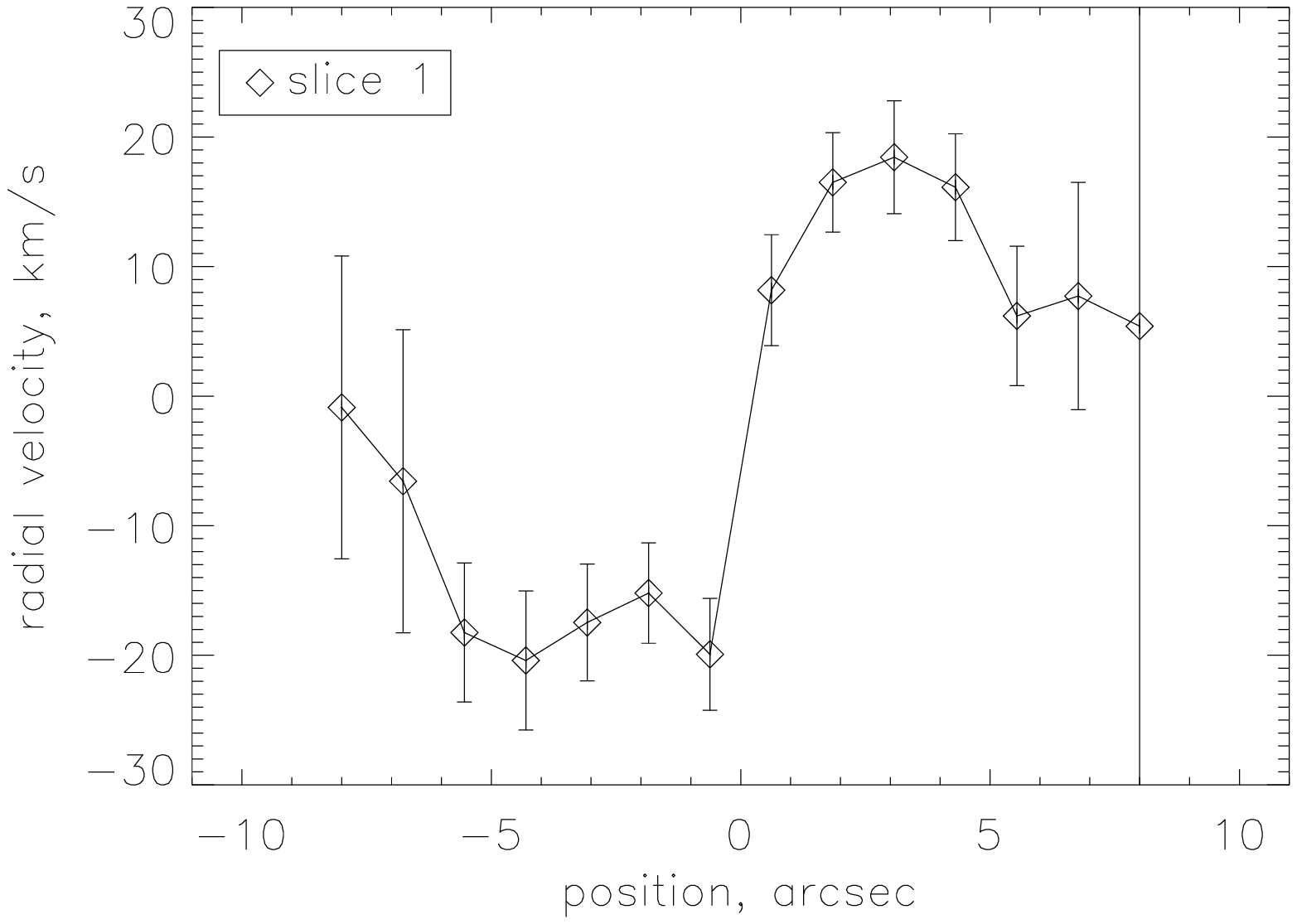} &
 \includegraphics[width=5.5cm]{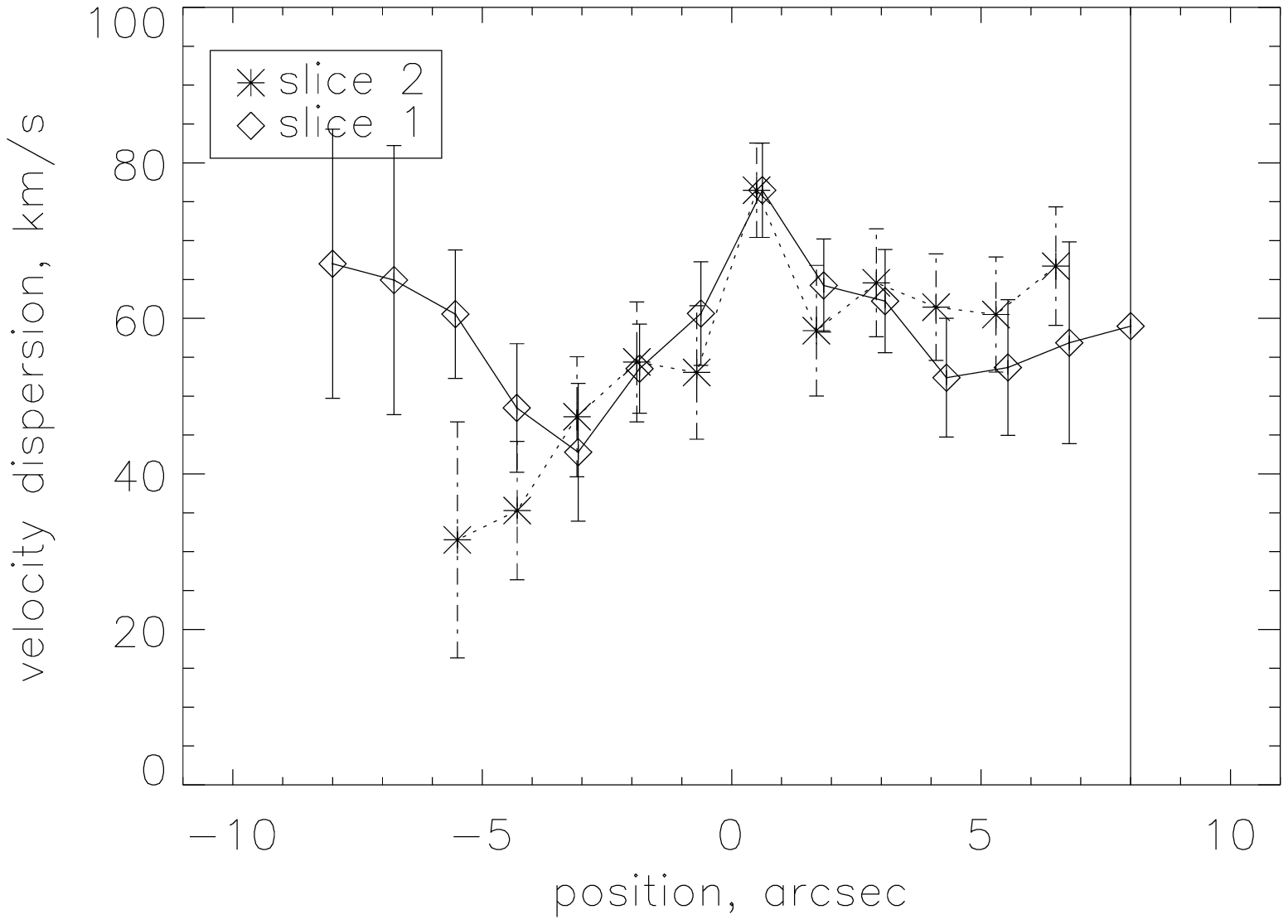} \\
 (a)&(b)\\
\end{tabular}
 \caption{(a)The radial velocity profile for the slice "1"\ (kinematical
major axis) and (b) velocity dispersion profiles for the slices "1"\ and "2"
represented on Fig.~1.}
\hfil
\label{prof}
\end{figure*}

\begin{figure*}
\hfil
\begin{tabular}{c c}
 \includegraphics[width=5.5cm]{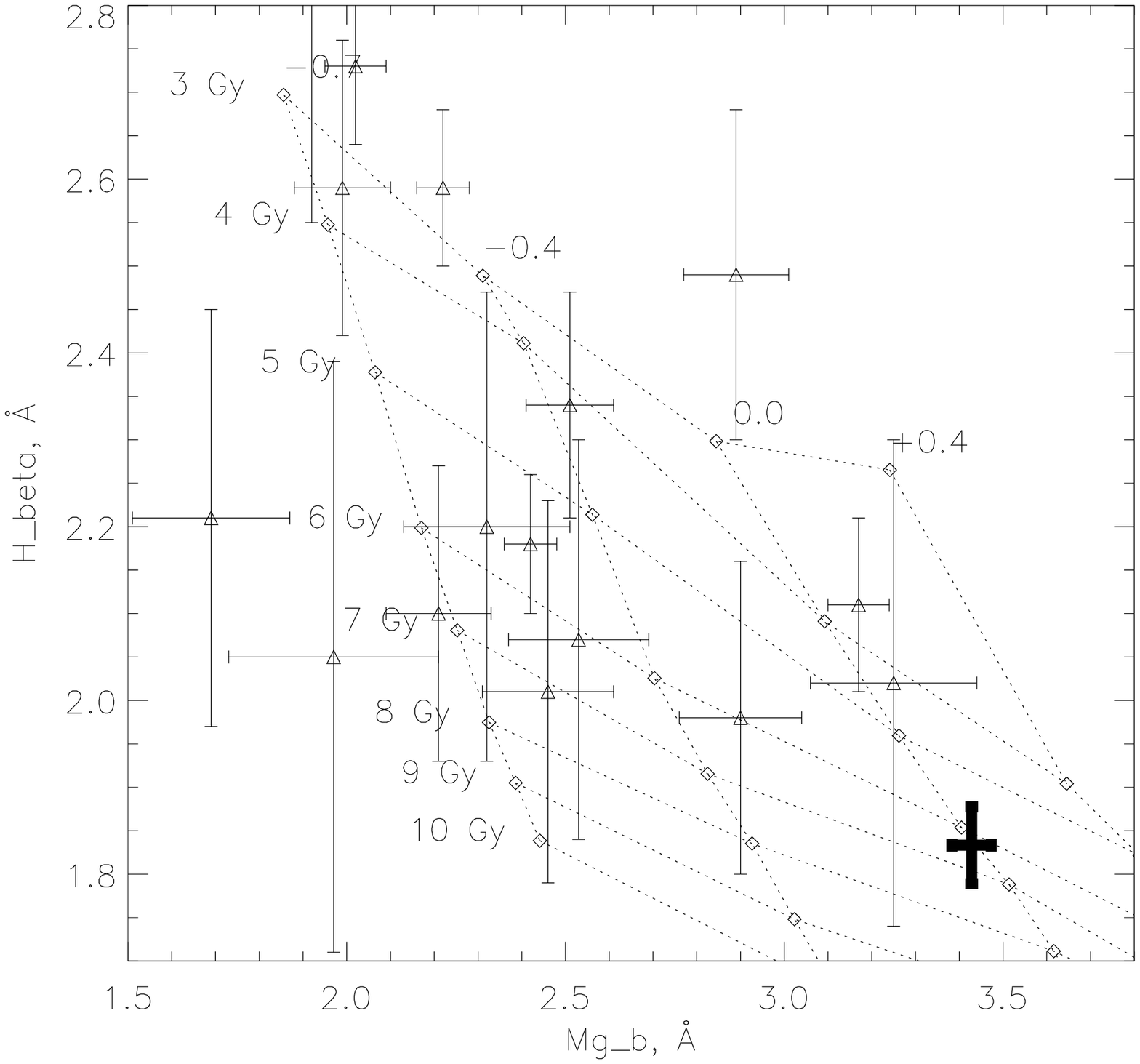} &
 \includegraphics[width=5.5cm]{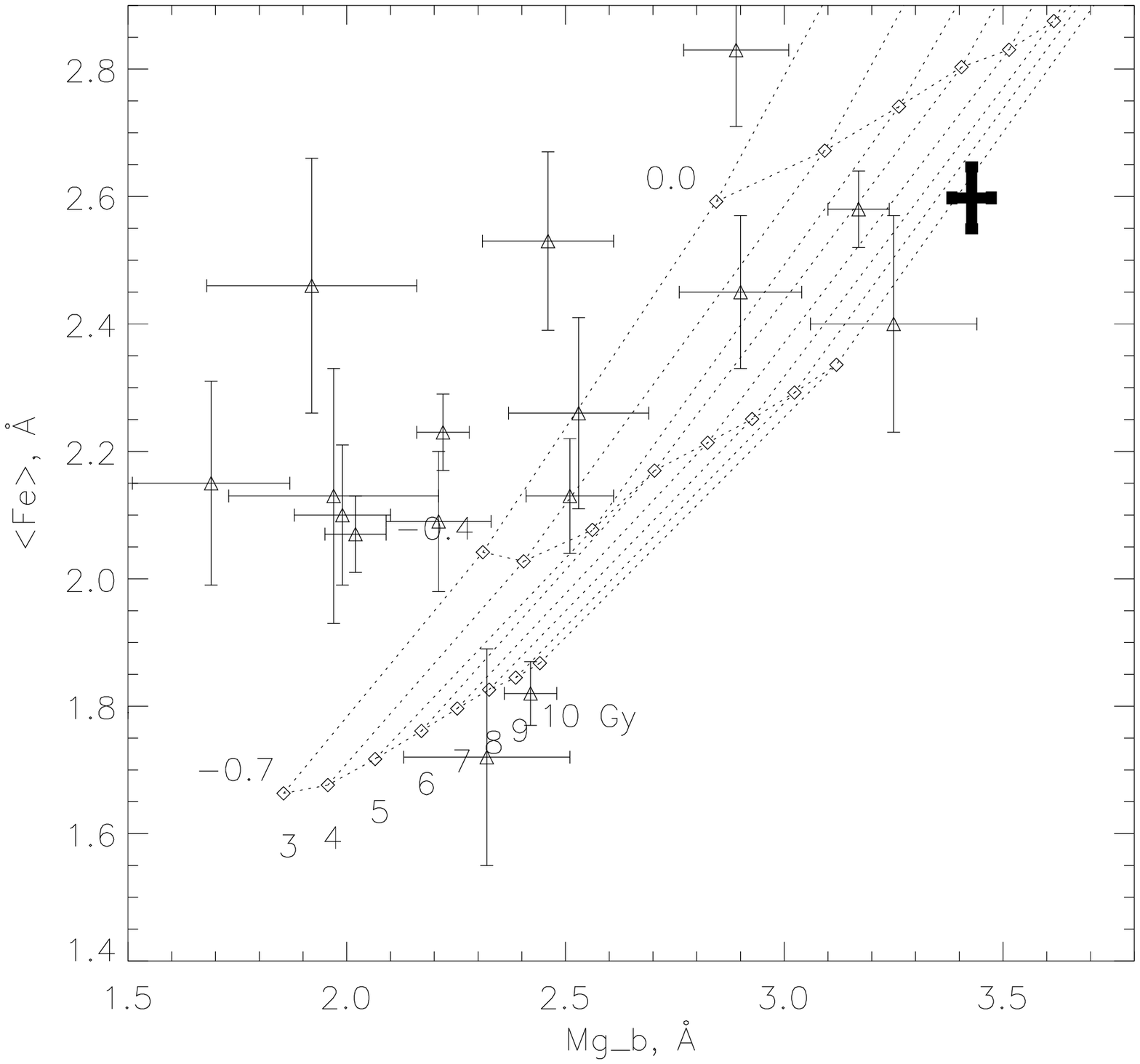} \\
 (a)&(b)\\
\end{tabular}
 \caption{Lick indices measurements of the stellar population properties
of IC3653 compared to the data for other dEs (Geha et al. 2003). Bold cross
represents our measurements for IC3653.}
\hfil
\label{lickidx}
\end{figure*}

\begin{table}
\caption{Comparison of main parameters of IC~3653 with NGC~205 and M~32.}
\label{table1}
\begin{tabular}{l | l l l l l l}
  & $M_B$ & $\mu_e$ & H$\beta$, \AA & Mg$b$, \AA & [Fe/H] & Age, Gy \\
\hline
  M32 & -15.5 & 18.7 & 2.20$\pm$0.06 & 2.97$\pm$0.06 & +0.05 & 4--5 \\
IC3653& -17.2 & 20.5 & 1.83$\pm$0.04 & 3.43$\pm$0.04 & -0.00 & 6.5$\pm$0.8 \\
NGC205& -15.7 & 22.5 & 4.20$\pm$0.23 & 0.77$\pm$0.25 & -0.37 & 0.9 \\
\end{tabular}
\end{table}

%------------ end of article ------------------->>

%% optional
%\section{Summary}

%% optional
\begin{acknowledgments}
We would like to thank the JENAM-2004 organizing committee for providing
financial support for one of the authors, CNRS for providing funding
bi-lateral collaboration, and SAO RAS time allocation committee for allocating
time on the 6-m telescope.
\end{acknowledgments}

%% appendix optional
%\appendix{This is the Appendix Title}
%This is an appendix with a title.

%\appendix{}
%This is an appendix without a title.

%
% Bibliography made with BibTeX:
%% kapalike is preferred if you have used \kluwerbib, above.
%% Otherwise you may use any .bst style your editor approves.

%%%%%%%%%%%%%%%%%%%%%%%%%%%%%%%
%This will allow many Bib\TeX\ bibliographies in one book.
%See the documentation, edbk.doc, for more information.
%
%\bibliographystyle{kapalike}
%\chapbblname{<name of .bbl file>}
%\chapbibliography{<name of .bib file>}
%
%%or 

%%%%%%%%%%%%%%%%%%%%%%%%%%%%%%%%

\begin{chapthebibliography}{99}

\bibitem[\protect\citeauthoryear{Barazza}{2000}]{2} 
Barazza F.D., Binggeli, B., Jerjen, H., 2002, A\&A, 391, 823

\bibitem[\protect\citeauthoryear{Conselice}{2001}]{19} 
Conselice C., Gallagher J. S. III, Wyse R. F. G., 2001, ApJ, 559, 791

\bibitem[\protect\citeauthoryear{Conselice}{2002}]{20} 
Conselice C., Gallagher J. S. III, Wyse R. F. G., 2002, AJ, 123, 2246

\bibitem[\protect\citeauthoryear{Conselice}{2003}]{18} 
Conselice C., O'Neil K., Gallagher J. S. III, Wyse R. F. G., 2003, ApJ, 591, 167

\bibitem[\protect\citeauthoryear{De Rijcke}{2001}]{7} 
De Rijcke S., Dejonghe H., Zeilinger W.W., Hau G., 2001, ApJ, 559, L21

\bibitem[\protect\citeauthoryear{De Rijcke}{2004}]{23} 
De Rijcke S., Dejonghe H., Zeilinger W. W., Hau, G. K. T., 2004, A\&A, 426, 53

\bibitem[\protect\citeauthoryear{Dekel}{1986}]{11} 
Dekel A., Silk J., 1986, ApJ, 303, 39

\bibitem[\protect\citeauthoryear{Duc}{2004}]{17} 
Duc P.-A., Bournaud F., Masset F., 2004, A\&A in press, astro-ph/0408524

\bibitem[\protect\citeauthoryear{Franx}{1988}]{21} 
Franx, M., Illingworth, J., 1988, ApJ, 327, L55

\bibitem[\protect\citeauthoryear{Geha}{2002}]{5} 
Geha M., Guhathakurta P., van der Marel R.P., 2002, AJ, 124, 3073

\bibitem[\protect\citeauthoryear{Geha}{2003}]{6} 
Geha M., Guhathakurta P., van der Marel R.P., 2003, AJ, 126, 1794

\bibitem[\protect\citeauthoryear{Jerjen}{2000}]{1} 
Jerjen H., Kalnajs A., Binggeli B., 2000, A\&A, 358, 845

\bibitem[\protect\citeauthoryear{Marcolini}{2003}]{15} 
Marcolini, A., Brighenti, F., D'Ercole, A., 2003, MNRAS, 345, 1329

\bibitem[\protect\citeauthoryear{Michielsen}{2003}]{9} 
Michielsen D., de Rijcke S., Dejonghe H., Zeilinger W. W., 2003, ApJ, 597, L21

\bibitem[\protect\citeauthoryear{Michielsen}{2004}]{10} 
Michielsen D., de Rijcke S., Zeilinger W. W., Prugniel Ph., Dejonghe H., Roberts, S., 2004, MNRAS, 311, 1293

\bibitem[\protect\citeauthoryear{Moore}{1998}]{16} 
Moore B., Lake G., Katz N., 1998, ApJ, 495, 139

\bibitem[\protect\citeauthoryear{Morelli}{2004}]{22} 
Morelli L., Halliday C., Corsini E. M., Pizzella A., Thomas D., Saglia R. P., Davies R. L., Bender R., Birkinshaw M., Bertola F., 2004, MNRAS, 354, 753

\bibitem[\protect\citeauthoryear{Mori}{1997}]{12} 
Mori M., Yoshii Y., Tsujimoto T., Nomoto K., 1997, 478, L21

\bibitem[\protect\citeauthoryear{Mori}{2000}]{13} 
Mori M., Burkert A., 2000, ApJ, 538, 559

\bibitem[\protect\citeauthoryear{Nagashima}{2004}]{14} 
Nagashima, M. \& Yoshii, Y., 2004, ApJ, 610, 23

\bibitem[\protect\citeauthoryear{Pedraz}{2002}]{4} 
Pedraz, S., Gorgas, J., Cardiel, N., S\'anchez-Bl\'azquez, P., Guzm\'an, R., 2002, MNRAS, 332, L59

\bibitem[\protect\citeauthoryear{Simien}{2003}]{24} 
Prugniel P., Simien F., 2003, ApSS, 284, 603

\bibitem[\protect\citeauthoryear{Simien}{2002}]{3} 
Simien F., Prugniel, Ph., 2002, A\&A, 384, 371

\bibitem[\protect\citeauthoryear{Simien}{2002}]{25} 
Trager S. C., Worthey G., Faber S. M., Burstein D., Gonzalez J., 1998, ApJS, 116, 1

\bibitem[\protect\citeauthoryear{van Zee}{2004}]{8} 
van Zee L., Skillman E.D., Haynes, M.P., 2004, AJ, 128, 121


\end{chapthebibliography}
\end{document}